\documentclass[11pt]{article}
\usepackage{graphicx}
\usepackage{float}
\title{\bf The Power Spectrum Of  Gravitational Waves in Anisotropic  Universe (\emph{Bianchi type--I})}
\author{Taimur Mohammadi,\footnote{University of Applied Science and Technology Kurdistan Branch , P.O.Box 66177-17792,
        Sanandaj, Iran. Email: t.mohammadi@uast.ac.ir }\\Behrooz Malekolkalami,\footnote{Department of Physics,
University of Kurdistan, P.O.Box 66177-15175,
        Sanandaj, Iran. Email: b.malakolkalami@uok.ac.ir}\\
       }
\begin{document}
\date{\today}
\maketitle
\begin{abstract}
One of the predictions from simple inflation models is  stochastic background of Gravitational Waves (\textbf{GW}), or literally what is called  Primordial Gravitational Waves (\textbf{PGW}) with a nearly scale--invariant  spectrum.  To discuss a possible direct detection of PGW, the quantity so--called Spectral Energy Density (\textbf{SED}) has crucial role. In this work, we consider PGW produced in the Radiation-Dominate(\textbf{RD}) era and generated by perturbing the isotropic and Anisotropic (\emph{Bianchi type--I}) metrics and focusing on the  SED  generated by these GW. This study was done because the power spectrum of GW from  the RD epoch is one of the most important topics in early cosmology, as GW produced during this period can provide us with direct information about very High Energies and fundamental phases of the universe (e.g. inflation and phase transition)The results   show that the power spectrums  in the early universe are diffetent for  isotropic and anisotropic   universe, but they coincide  at the present time.
\end{abstract}

\medskip
{\small \noindent
\hspace{5mm}

\bigskip

\section{Introduction}
\indent
One of the central predictions of Einstein's general theory of relativity is that GW will be generated by accelerating masses~\cite{GWs 1916, GWs 1918}. For decades, it was so difficult to analyze and to define the energy and momentum carried by GW. The first direct observation from merging massive black holes reported on September 14, 2015~\cite{Abbott 2016}. This has become a renewed attention to detect new information in astronomy and cosmology. The GW are very important physical process that can be studied to give us valuable information about the dynamics of spacetime geometry and to probe in the history of the Universe.
Many researches have been done in this field, among which the follows can be mentioned, e. g. ~\cite{Abbott 20171}--\cite{Michele 2012}.\\
Nowadays, the sources of GW are largely known including gravitational collapse, coalescing binaries, pulsars, rapidly spinning accreting neutron stars, and the stochastic background.  the Primordial Gravitational Waves (PGW) produced in the early Universe are  a type of stochastic background which emanate from regions of strong gravity.
They also carry uncorrupted physical signatures of early universe and its structure. These waves form an extremely large number of weak, independent, and unresolved sources and this makes the waves have a random character. At the present time, GW originated from  inflationary period will be out of reach for all planned ground--based instruments. The detection of such a background would have a profound impact on early Universe cosmology and other research fields of physics such as High Energy physics which includes  high energies events, that will never be accessible by other means.\\
One of the useful tools  for  studying and analyzing the  stochastic waves  is the spectral method that we apply according to its random nature, then
the main reason to use the spectral method is due to the random nature of the stochastic waves. Generally, for a periodic signal,  knowing the spectrum frequency and power of each harmonic contributing to the signal allows it to be decomposed into its component parts. Hence, one reconstructs  the signal from its component, in which case the signal becomes more meaningful phase data. The PGW  carry important information from the earliest phases of  the universe  and due to the various Astrophysical sources contributions,  the information is disturbed. One of the important quantities in spectral methods to extract such information is the power spectrum, or more specifically in our discussion, known as Spectral Energy Density (SED) quantity, denoted by  $\Omega_{GW}(f)$. It is hoped that such  method and tool can be of great help to provide a snapshot of the early universe. Also, It  can help us to have a relative scheme of energy scale of the early universe, due to the Inflation Energy Scale is one of the main challenges in theoretical and experimental High Energy physics.\\
The study of PGW with spectral considerations in isotropic background has already been done in the works, e. g.  \cite{Yuki 2006, Latham 2005, Ryusuke 2014}. Here, we do this for the case in which the background is  an anisotropic metric (\emph{Bianchi type--I}) in  the Einstein's gravity. The study of GW from the RD period is of great important in cosmology, High Energy physics, and theories of gravity. This waves act as transparent messenger from the early universe and can provide us with information that is not available in any other way (e.g. CMB photons).\\
Our approach to obtain evolution equations is the Lagrangian method.  As we will see, to obtain the  evolution equations, using the scale factor instead of conformal--time would be a good trick to avoid the piece--wise function ( as used by the previous  works,  e. g. [14, 15, 16,17, 18], in tha isotropic case).\\
The work is organized as follows:
In section II, we introduce the motion equations of tensor perturbations essential for the work. In section III,  the power spectrum of perturbations in Einstein's framework is investigated.  Section IV points out future works and the conclusions are given in section V.
\section{Evolution Equations}
In this section, we  introduce the equation of motion  for the  (tensorial) perturbation of the anisotropic (\emph{Bianchi type--I}) background metric $\bar{g}_{\mu\nu}$ given by
\begin{equation}\label{N7}
ds^2=\bar{g}_{\mu\nu}dx^{\mu}dx^{\nu} =-dt^{2}+ A^{2}dx^{2}+B^{2}dy^{2}+C^{2}dz^{2},
\end{equation}
where $A(t), B(t), C(t)$ are  time--dependent scale factors.  The energy momentum tensor is given by
\begin{equation}\label{N7}
T_{ij}=(\rho+p)v_{i}v_{j}+pg_{ij},
\end{equation}
where $v_{i}$ is the four velocity vector and  $\rho$, $p$ are energy density and thermodynamic pressure,   satisfying the perfect fluid form, that is
\begin{equation}\label{N7}
p=\omega\rho, \hspace{5mm} 0\leq\omega\leq1.
\end{equation}
\begin{equation}\label{N7}
R_{ij}-\frac{1}{2}Rg_{ij}=-T_{ij}+\Lambda g_{ij},
\end{equation}
where $\Lambda$ is the cosmological constant  and equal to $\Lambda=\beta H^{2}$ \cite{Mukesh 2016} (where $\beta$ is the positive constant and H is the Hubble parameter. By substituting equation (1) into equation (4), $A(t), B(t)$ and $C(t)$ are obtained as follows  \cite{Mukesh 2016}
\begin{equation}\label{N7}
B(t) = \left((3-\beta)(c_{1}t+c_{2})\right)^{\frac{1}{3-\beta}}\exp \left(\frac{k_{2}-k_{1}}{3((3-\beta)(c_{1}t+c_{2}))^{\frac{3}{3-\beta}}}\right)=a(t)\exp \left(\frac{k_{2}-k_{1}}{3a^{3}(t)}\right)
\end{equation}
\begin{equation}\label{N7}
A(t) = \left((3-\beta)(c_{1}t+c_{2})\right)^{\frac{1}{3-\beta}}\exp \left(\frac{2k_{1}+k_{2}}{6((3-\beta)(c_{1}t+c_{2}))^{\frac{3}{3-\beta}}}\right)=\sqrt{a(t) B(t)}\exp\left(\frac{k_{1}}{2a^{3}(t)}\right)
\end{equation}
\begin{equation}\label{N7}
C(t) = \left((3-\beta)(c_{1}t+c_{2})\right)^{\frac{1}{3-\beta}}\exp\left(\frac{2k_{2}-k_{1}}{2((3-\beta)(c_{1}t+c_{2}))^{\frac{3}{3-\beta}}}\right)=\sqrt{\frac{B^{3}(t)}{a(t)}}\exp\left(\frac{k_{2}}{2a^{3}(t)}\right)
\end{equation}
where $c_{i}$ and $k_{i} $ are integration constants. Also,  $\beta$ is the positive constant and $a(t)$ is an average scale factor, defined as
\begin{equation}\label{N7}
a(t) = V^{\frac{1}{3}}(t)=\left(A(t)B(t)C(t)\right)^{\frac{1}{3}}=\left((3-\beta)(c_{1}t+c_{2})\right)^{\frac{1}{3-\beta}},
\end{equation}
where $V(t)$ is spatial volume. \\
We begin by disturbing the metric (1) as follows:
\begin{equation}\label{N7}
ds'^2=g_{\mu\nu}dx^{\mu}dx^{\nu}=\left(\bar{g}_{\mu\nu}+h_{\mu\nu}\right)dx^{\mu}dx^{\nu} =\left(- a^{2}d\tau^{2}+A^{2}dx^{2}+B^{2}dy^{2}+C^{2}dz^{2}\right)+a^{2}h_{ij}dx^{i}dx^{j},
\end{equation}
where  $h_{ij}=h_{ij}(\tau,\textbf{x})$ are the perturbation potentials  satisfying, symmetric ($h_{ij}$=$h_{ji}$), traceless ($h^i_{i} = 0)$ and transverse $(h^j_{i,j} = 0)$ conditions.\\
The equations of motion for perturbations, in a general $f(R)$ gravity, are obtained by variation of the following action
\begin{equation}\label{C00}
S =\int d\tau d\textbf{x}\sqrt{-\bar{g}}\left(f(R)+\frac{1}{2}\Pi_{ij}h_{ij}\right),
\end{equation}
where $\Pi_{ij}$ is the anisotropic stress  tensor\cite{Steven 2003}.
For the isotropic perturbations (that is $h_{ij}(\tau,x,y,z)=h(\tau,x,y,z)= h(\tau,\textbf{x})$) and  the vacuum or  perfect fluid ($\Pi_{ij} =  0$) cases, the equations of motion take the following form~\cite{Taimur 2020, Taimur 2024}
\begin{equation}\label{C00}
\partial_{\mu}\left(\sqrt{|\bar{g}|}\hspace{1mm} f'(R) \hspace{1mm} \frac{\partial R}{\partial(\partial _{\mu}h)}\right)=0.
\end{equation}
where prime denotes  derivative respect to $R$. For the Einstein's gravity  $f(R)=R$ and the $R$ corresponding to perturbed metric (9) is  given by~\cite{Latham 2005}
\begin{equation}\label{C00}
R=-\frac{\bar{g}^{\mu\nu}}{64 \pi G}\left(\partial_{\mu}h \partial_{\nu}h\right).
\end{equation}
By inserting these prsented above  into (11), we get:
\begin{equation}\label{C00}
\partial_{\mu}\left(\sqrt{|\bar{g}|}\hspace{1mm}  \hspace{1mm}\bar{g}^{\mu\nu}\partial _{\nu}h\right)=0.
\end{equation}
In the next section, we use the latter equation to obtain the evolution of perturbations.
\section{The Evolution  Equations}
By substituting the anisotropic  background metric (9) into the equation (13), the evolution equation for the metric perturbations reads
\begin{equation}\label{C00}
h^{''}+2\left(\frac{a^{'}}{a}\right) h^{'}+a^{2}\left(\frac{\partial_{x}^{2}}{A^{2}}+\frac{\partial_{y}^{2}}{B^{2}}+\frac{\partial_{z}^{2}}{C^{2}}\right)h=0,
\end{equation}
where $h=h(\tau, \textbf{x})$ and prime is derivative respect to the conformal time  $\tau$.\\
By taking Fourier transforms of both sides (14), one gets
\begin{equation}\label{C00}
h''(\tau, \textbf{k})+2\left(\frac{a^{'}}{a}\right) h'(\tau, \textbf{k})+(\frac{ka}{\sqrt{3}})^{2}\left(A^{-2}+B^{-2}+C^{-2}\right)h(\tau, \textbf{k})=0,
\end{equation}
where $h(\tau, \textbf{k})$ denotes  the Fourier transform of $h(\tau, \textbf{x})$ and $k=\sqrt{k_{x}^{2}+k_{y}^{2}+k_{z}^{2}}$  is the wave number (with $k_{x}=k_{y}=k_{z}=\frac{k}{\sqrt{3}}$ ).

As mentioned  above,   the prime is derivative respect to the conformal time  $\tau$. Also, note that, in equations (14) and (15), the directional scale factors $A, B, C$ are explicit functions of average scale factor $a$. On the other hand,   the relation  $dt = a d\tau$ allows us to replace the derivatives respect to the $\tau$ with respect to $a$.  With this replacement,  the perturbation becomes a function of $a$ (that is  $h(\tau, \textbf{x})\rightarrow h(a, \textbf{x})$, or equivalently $h(\tau, \textbf{k})\rightarrow h(a, \textbf{k})$) Therefore, equation (15) takes the following form \cite{Taimur 2020}:
\begin{equation}\label{C00}
a^{4}H^{2}h_{aa}(a, k) + \left(4 a^{3}H^{2}+\frac{a^{4}}{2}\frac{dH^{2}}{da}\right)h_{a}(a, k)+(\frac{ka}{\sqrt{3}})^{2}\left(A^{-2}+B^{-2}+C^{-2}\right)h(a, k)=0,
\end{equation}
where  $h_{a}(a, k)=\frac{dh(a, k)}{da}$ ,  $h_{aa}(a, k)=\frac{d^{2}h(a, k)}{da^{2}}$ and $H(=\frac{\dot{a}}{a}=h'/h_{a}$) is the Hubble parameter. Now, knowing the Hubble parameter as a function of scale factor $H=H(a)$,  equation (16) becomes uniform, meaning that it consists of  one dependent variable $a$ and one independent variable $h(a,k)$. It is not difficult to show that here the Hubble parameter has the same form in the isotropic case,  that is
\begin{equation}\label{C00}
H^{2}(a)= H^{2}_{0}\left(\Omega_{r}a^{-4}+\Omega_{m}a^{-3}+\Omega_{\Lambda}\right),
\end{equation}
where $\Omega_{r}=9.4\times10^{-5}\simeq 10^{-4}$, $\Omega_{m}=0.3$, and  $\Omega_{\Lambda}=0.7$ are radiation, matter and  dark energy density parameters respectively, and $H_0\simeq (72\pm8) km s^{-1} Mpc^{-1}\simeq(2.3\pm0.3)\times10^{-18} s^{-1}$ is the Hubble constant. In this work, we investigate PGW emitted from the dominate radiation epoch, then
\begin{equation}\label{C00}
H^{2}(a)= H^{2}_{0}\Omega_{r}a^{-4},
\end{equation}
by replacing equation (18) into equation (16), we get
\begin{equation}\label{C00}
h_{aa}(a, k) +\frac{2}{a}h_{a}(a, k)+(\frac{100ka}{\sqrt{3}H_{0}})^{2}\left(A^{-2}+B^{-2}+C^{-2}\right)h(a, k)=0.
\end{equation}
The latter equation plays a central role in our subsequent analysis and discussion, and since the solutions to this equation depend on the values  of the parameters $k_{1}$ and $k_{2}$ in equations (5-7), we will consider these solutions for selected values of the parameters in the following.
\subsubsection{$k_{1}= k_{2}=0$}
In this case $A=B=C=a$ and equation (19) becomes
\begin{equation}\label{C00}
h_{aa}(a, k) +\frac{2}{a}h_{a}(a, k)+(\frac{100k}{H_{0}})^{2}h(a, k)=0.
\end{equation}
As expected, this equation corresponds to the isotropic case  that  has been investigated in details in previous work \cite{Taimur 2020}. Since the scale factor was very small during the radiation era ($a (t)\ll1$), equation (20) can be written as follows, taking into account the new variable  $\chi=\frac{1}{a}$:
\begin{equation}\label{C00}
\chi^{4}\frac{\partial^{2} h}{\partial \chi^{2}}+(\frac{100k}{H_{0}})^{2}h(\chi, k)=0.
\end{equation}
To obtain the latter equation, we use the following relations:\\
$h_{a}=\frac{\partial h}{\partial a}=\frac{\partial h}{\partial \chi}\frac{\partial \chi}{\partial a}=-\chi^{2}\frac{\partial h}{\partial \chi}, h_{aa}=\frac{\partial^{2} h}{\partial a^{2}}=\frac{\partial }{\partial a}(-\chi^{2}\frac{\partial h}{\partial \chi})=2\chi^{3}\frac{\partial h}{\partial \chi}+ \chi^{4}\frac{\partial^{2} h}{\partial \chi^{2}},$\\
where the only kept the fourth order terms of  $\chi$ due to  $\chi=\frac{1}{a}\gg1$.\\
It isn't difficult to show  that  the equation (21) has the following  general solution:
\begin{equation}\label{N7}
h(a,k)=\frac{b_{1}}{a}\exp(\frac{100 i a k}{ H_{0}})-\frac{ib_{2} H_{0}}{200 a k}\exp(\frac{-100 i a k}{ H_{0}}),
\end{equation}
where $b_i$ are integration constant which  by imposing  the following initial conditions:
\begin{eqnarray}
h\left(a_{r}, k\right)=1,  \\ \hspace{2.5mm} \nonumber
h_{a}\left(a_{r}, k\right)= 0,\\ \nonumber
a_{r}\simeq 10^{-5},
\end{eqnarray}
are obtained as
$b_{1}=\frac{(k-1000 i H_{0})\exp(\frac{-ik}{1000H_{0}})}{200000k}$ and $b_{2}=\frac{(ik-1000  H_{0})\exp(\frac{ik}{1000H_{0}})}{1000H_{0}}.$
\subsubsection{$k_{1}= k_{2}=\alpha$}
In this case, the directional scale factors becomes: $B=a$ and $A=C=a\exp (\frac{\alpha}{2a^{3}})$. By substituting these  into equation (19), we get
\begin{equation}\label{C00}
h_{aa}(a, k) +\frac{2}{a}h_{a}(a, k)+(\frac{100k}{\sqrt{3}H_{0}})^{2}\left(1+2\exp(-\frac{\alpha}{a^{3}})\right)h(a, k)=0,
\end{equation}
which by new variable $\chi= 1/a$ reads
\begin{equation}\label{C00}
\chi^{4}\frac{\partial^{2} h(\chi, k)}{\partial \chi^{2}} +(\frac{100k}{\sqrt{3}H_{0}})^{2}\left(1+2\exp(-\alpha\chi^{3})\right)h(\chi, k)=0.
\end{equation}
Because $\chi\gg1$, if $\alpha$ is such that $\exp(-\alpha\chi^{3})\simeq 1-\alpha \chi^{3}$, so equation (25) become
\begin{equation}\label{C00}
\chi^{4}\frac{\partial^{2} h(\chi, k)}{\partial \chi^{2}} +(\frac{100k}{\sqrt{3}H_{0}})^{2}\left(3-2\alpha \chi^{3}\right)h(\chi, k)=0,
\end{equation}
In other words (because of $\alpha\chi^{3}\gg 1$, the first part in parentheses can be omitted compared to $2\alpha\chi^{3}$  )
\begin{equation}\label{C00}
\chi\frac{\partial^{2} h(\chi, k)}{\partial \chi^{2}} -2\alpha(\frac{100k}{\sqrt{3}H_{0}})^{2} h(\chi, k)=0,
\end{equation}
This equation has solution as follows
\begin{equation}\label{N7}
h(\chi,k)=-k\sqrt{\chi \lambda}BesselI(1,2k\sqrt{\chi \lambda})b_{1}+2k\sqrt{\chi \lambda}BesselK(1,2k\sqrt{\chi \lambda})b_{2},
\end{equation}
where $\lambda=2\alpha(\frac{100}{\sqrt{3}H_{0}})^{2}$, $b_{i}$  are integration constants and are obtained from the initial conditions consider in equation (22), $BesselI$ and $BesselK$ are Bessel functions of type I and type K. This equation in terms of $a$ will be as follows
 \begin{equation}\label{N7}
h(a,k)=-k\sqrt{\frac{\lambda}{a} }BesselI(1,2k\sqrt{\frac{\lambda}{a} })b_{1}+2k\sqrt{\frac{\lambda}{a} }BesselK(1,2k\sqrt{\frac{\lambda}{a} })b_{2},
\end{equation}
\subsubsection{$k_{1}\neq k_{2} $ and $k_{1}=\epsilon_{1},  k_{2}=\epsilon_{2}$}
In this case $B=a \exp \left(\frac{\epsilon_{2}-\epsilon_{1}}{3a^{3}}\right), A= a\exp \left(\frac{\epsilon_{1}}{2a^{3}}\right)\sqrt{\exp \left(\frac{\epsilon_{2}-\epsilon_{1}}{3a^{3}}\right)}$ and $C= a\exp \left(\frac{\epsilon_{2}}{2a^{3}}\right)\sqrt{\exp \left(\frac{\epsilon_{2}-\epsilon_{1}}{a^{3}}\right)}$ and equation (19) becomes
\begin{equation}\label{C00}
h_{aa}(a, k) +\frac{2}{a}h_{a}(a, k)+(\frac{100k}{\sqrt{3}H_{0}})^{2}\left(\exp({\frac{-2\left(\epsilon_{1}+3\epsilon_{2}\right)}{3a^{3}})}\left(\exp({\frac{5\epsilon_{1}}{3a^{3}})}+\exp({\frac{5\epsilon_{2}}{3a^{3}})}+\exp({\frac{4(\epsilon_{1}+\epsilon_{2})}{3a^{3}})}\right)\right)h(a, k)=0,
\end{equation}
By replacing the new variable $\chi=\frac{1}{a}$ and considering the explanations in the previous section, this equation is rewritten as follows (for example $\epsilon_{1}=10^{-10}$ and $\epsilon_{2}=10^{-12}$)
\begin{equation}\label{C00}
\chi^{4}\frac{\partial^{2} h(\chi, k)}{\partial \chi^{2}} +(\frac{100k}{\sqrt{3}H_{0}})^{2}\left(3+\frac{97\chi^{3}}{10^{12}}\right)h(\chi, k)=0.
\end{equation}
considering that  $\frac{97\chi^{3}}{10^{12}}\gg3$, this equation will be as follows
\begin{equation}\label{C00}
\chi\frac{\partial^{2} h(\chi, k)}{\partial \chi^{2}} +\left(1.2436\times10^{11}\right)k^{2}h(\chi, k)=0.
\end{equation}
\subsection{The Power Spectrum}
The main property of  a stochastic background of GWs of cosmological origin is frequency spectrum and one of the useful characterization of the spectrum is SED. To characterize this, we need to introduce  the spectral amplitude $\Delta_{h}^{2}(\tau,k)$ defined by:
\begin{equation}\label{N7}
<h_{ij}(\tau,\textbf{x})h^{ij}(\tau,\textbf{x})>=\int\frac{dk}{k}\Delta_{h}^{2}(\tau,k),
\end{equation}
which can be written in the reverse form as
\begin{equation}\label{N7}
\Delta_{h}^{2}(\tau,k)=\frac{k^{3}}{\pi^{2}} <|h(\tau, k)|^{2}>,
\end{equation}
or, in terms of scale factor
\begin{equation}\label{N7}
\Delta_{h}^{2}(a,k)=\frac{k^{3}}{\pi^{2}} <|h(a, k)|^{2}>.
\end{equation}
This amplitude relates the spectral distribution of the amplified fluctuations and
the cosmological kinematic parameters. It  is also useful to describe the distribution of the  modes in outside the horizon.\\
It is quite common, when disscusing  the subject of the GWs detection, to define the dimensionless parameter (SED) as
\begin{equation}\label{N7}
\Omega_{h}(a,k)=\frac{1}{\rho_c}\frac{d\rho}{d\ln k },
\end{equation}
where $\rho$  and $\rho_c$ are   GW energy density and critical energy density, respectively. Since, the relic GW  with mode inside horizon should be still present today,  they must be accessible to direct observations. The SED characterises the spectrum of the relic waves and thus it is useful to discuss a possible their direct detection. For the mode inside horizon, the SED is related to the spectral amplitude through the relation
\begin{equation}\label{N7}
\Omega_{h}(a,k)=k^2\frac{\Delta_{h}^{2}}{12\hspace{0.5mm}a^2H^{2}(a)}=\frac{k^5}{12\pi^{2}}\frac{<|h(a, k)|^{2}>}{\hspace{0.5mm}a^2H^{2}(a)}.
\end{equation}
The spectrum of waves at the present time $\tau_0$ is obtained by  substituting the conventional value of scale factor ($a(\tau_0)=a_0=1$) into (34), leading to
\begin{equation}\label{N7}
\Omega(k)=\Omega_{h}(1, k)=\frac{k^5}{12\pi^{2}}\frac{<|h(1, k)|^{2}>}{\hspace{0.5mm}H_0^{2}}=\frac{k^5}{3\pi^{2}}\frac{|h(k)|^{2}}{\hspace{0.5mm}H_0^{2}},
\end{equation}
where  we have used $|h(k)|^{2}=\frac{1}{2}\left(\langle |h_{+}|^{2}\rangle+\langle |h_{\times}|^{2}\rangle\right)=\frac{1}{4}\langle h^{ij} h_{ij}\rangle=\frac{1}{4}<|h(1, k)|^{2}>$\rlap.\footnote{The relations mean that the contributions of the two GW polarization cases ($+,\times$) are taken to be equal \cite{Yuki 2006}.}\\
Equation (34) or (35) represents  form of the spectrum  dependence on the perturbations which must be determined from equations (23)and (29), we have calculated the SED for isotropic and anisotropic models and results are shown in Figures 1 and  2 respectively.\\
A comparison of the two isotropic and anisotropic models is shown in Figure 3. It can be seen that over time and up to the present, both two models are coincident and compatible with each other.  This will be very important result for researches and shows that regardless of whether the early universe is considered isotropic or anisotropic, theoretical and observational results with give us an isotropic universe at the present time (age of universe).
\begin{figure}[H]
\centering
\includegraphics[width=0.6\textwidth]{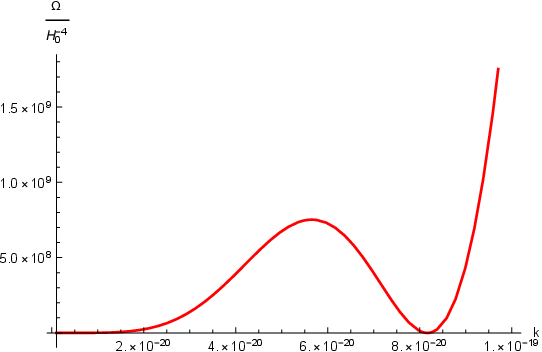}
\caption{The power spectrum in isotropic model ($k_{1}=k_{2}=0$).}
\label{5bl0tn}
\end{figure}
\begin{figure}[H]
\centering
\includegraphics[width=1.1\textwidth, height=0.2\textheight]{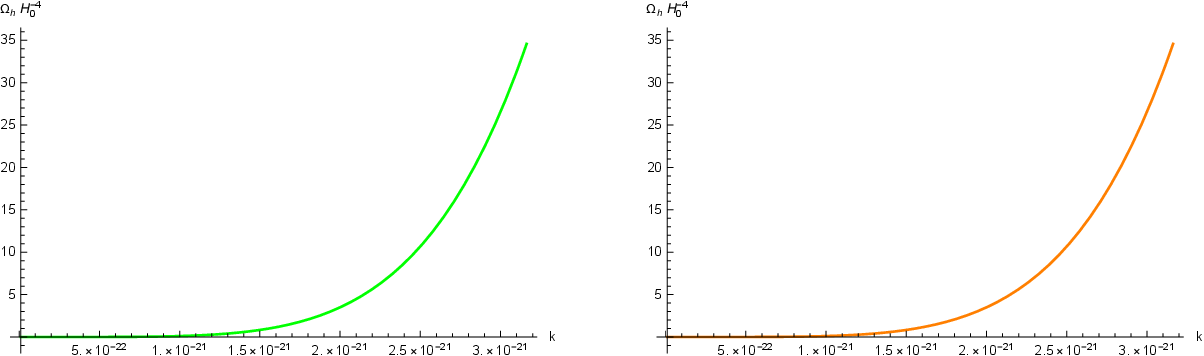}
\caption{The power spectrum in  anisotropic model (Green for $k_{1}=k_{2}=\alpha=10^{-20}$ and Orange for $k_{1}\neq k_{2} $ and $k_{1}=\epsilon_{1} =10^{-10},  k_{2}=\epsilon_{2}=10^{-12}$ ).}
\label{5bl0tn}
\end{figure}
\begin{figure}[H]
\centering
\includegraphics[width=1.1\textwidth, height=0.2\textheight]{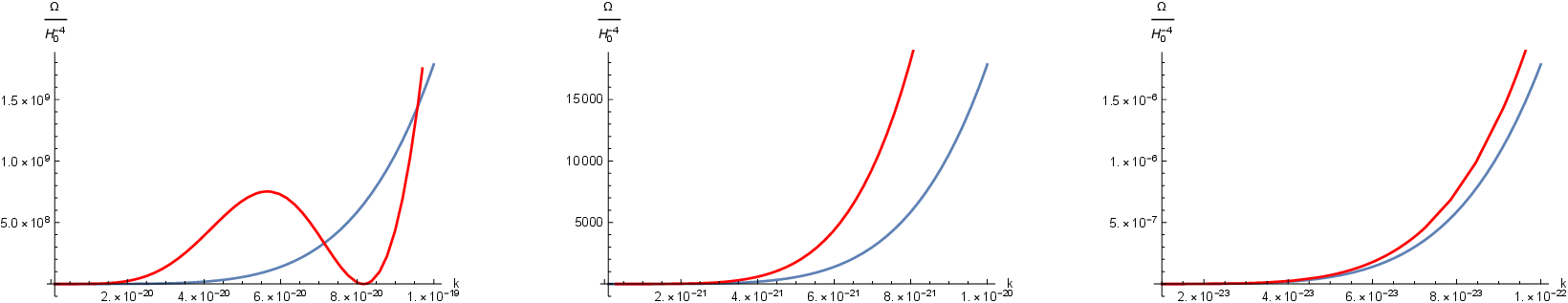}
\caption{The power spectrum in isotropic and anisotropic models (Blue for anisotropic model with $k_{1}=k_{2}=\alpha=10^{-25}$, and Red for isotropic model with $k_{1}=k_{2}=0$).}
\end{figure}
The figures show that the models are consistent at now (age of universe). Therefor in cosmological models, especially on large scales and as the universe approaches its current state, the behaviors of space-time become very close to each other. This indicates the convergence of the properties of space-time to an isotropic state. For example, in large-scale cosmological physics, sometimes the effects of anisotropy in space-time can disappear and the energy distribution behaves isotropically. this can happen especially at recent times and large distances, where GW or cosmic microwave background radiation (CMB) can be observed to resemble an isotropic distribution more than they are affected by anisotropy.\\
The fact that the spectra have converged at the present time could indicate that on the current time scale, the differences between isotropic models are becoming less pronounced. In other words, over time, the properties of space-time may naturally tend towards an isotropic behavior, a phenomenon that can be explained by the effects of homogenization in cosmology and different cosmological models.\\
In generally if the spectra are perfectly aligned at the present time ($\emph{f}=\frac{ kc}{2\pi}\simeq 10^{-15} Hz $), it can be interpreted that under current conditions, the effects of anisotropy in space-time have weakened to such an extent that isotropic and anisotropic models exhibit almost identical behavior.\\
Finally, experimental observations could show that data collected from GW (e.g. LIGO or Virgo) or cosmic microwave background (CMB)  radiation at the present time are such that the differences predicated by isotropic and anisotropic models are indistinguishable.
\section{Future Work}
The field of future work could be the study of GWs in other anisotropic universe such as Bianchi type II, III and else. Also, investigate $f(R)$ gravity models in anisotropic universes.
\section{Conclusions}
The power spectrum  corresponding to the primordial metric perturbations (leads to PGW) arisen from  anisotropic background (\emph{Bianchi type--I metric}) are presented in the Einstein's  gravities. The main conclusions are summarized as follows:\\
1) Despite the fact that,  we consider the early universe to be isotropic or anisotropic, the power spectrum of the GW detected at the present time (age of universe) will be the same, although they are different at the first (early universe).\\
2) We consider an average scale factor for all three directions, which is in full agreement with the results of isotropic models at present.\\
3) If the spectra coincided at the present time (age of universe), this would represent an interesting result in cosmological and gravitational physics analysis. This agreement could particulary emphasize the fact on large scales and in recent times, the universe may have approached an isotropic state. This could have interesting consequences in various fields of cosmology, gravitational physics and GW modeling.

\end{document}